# Role of quantum fluctuations in structural dynamics of liquids of light molecules

A. Agapov[1,*], V.N. Novikov[1], A. Kisliuk[2], R. Richert[3], A.P. Sokolov[1,2,4]

[1]*Department of Chemistry and Joint Institute for Neutron Sciences, University of Tennessee, Knoxville, TN 37996, USA*

[2]*Chemical Sciences Division, Oak Ridge National Laboratory, Oak Ridge, TN 37831, USA*

[3]*School of Molecular Sciences, Arizona State University, Tempe, AZ 85287-1604*

[4]*Department of Physics and Astronomy, University of Tennessee, Knoxville, TN 37996*

## Abstract

A possible role of quantum effects, such as tunneling and zero-point energy, in the structural dynamics of supercooled liquids is studied by dielectric spectroscopy. Presented results demonstrate that the liquids, bulk 3-methyl pentane (3MP) and confined normal and deuterated water have low glass transition temperature and unusually low for their class of materials steepness of the temperature dependence of structural relaxation (fragility). Although we don't find any signs of tunneling in structural relaxation of these liquids, their unusually low fragility can be well described by the influence of the quantum fluctuations. Confined water presents especially interesting case in comparison to the earlier data on bulk low-density amorphous and vapor deposited water. Confined water exhibits much weaker isotope effect than bulk water, although the effect is still significant. We show that it can be ascribed to the change of the energy barrier for relaxation due to a decrease in the zero-point energy upon D/H substitution. The observed difference in the behavior of confined and bulk water demonstrates high sensitivity of quantum effects to the barrier heights and structure of water. Moreover, these results demonstrate that extrapolation of confined water properties to the bulk water behavior is questionable.

## I. Introduction

Quantum effects, including tunneling, are well documented in glasses at very low temperatures, T<1-10K [1-5]. There have been many studies demonstrating tunneling effects in thermal conductivity, specific heat, acoustic attenuation, and dielectric studies [3,4,5]. But all these studies were focused on very low temperatures, T<10K. However, several recent papers raised a question of possible quantum effects in dynamics and glass transition of light molecules even at much higher temperatures [6-13]. One "trivial" quantum effect arises due to zero-point quantum fluctuations. For light atoms zero-point vibrations lead to a large amplitude of quantum fluctuations, that can be measured through the mean-squared displacements MSD $<u_0^2>$, and to a rather high zero-point energy, i.e. the lowest energy level in the potential well that effectively decreases the energy barrier for structural rearrangements [3]. Another effect, quantum tunneling, depends exponentially on atomic (molecular) mass and its probability significantly increases for light molecules [3]. Since the glass transition by definition corresponds to structural relaxation time $\tau_\alpha$ of the order of 100 seconds, very long in comparison to microscopic vibrational times, the probability of tunneling on these long times might be appreciable for light molecules even at relatively high temperatures, e.g., at T>100K in water [8,9,10]. In particular, recent analysis suggested [8] that if the glass transition temperature $T_g$ is lower than the half of the Debye temperature $T_D$, quantum effects might not be negligible at $T_g$. They will broaden the glass transition range pushing $T_g$ below the traditional $2/3T_m$ value ($T_m$ is melting temperature), and might lead to a sub-Arrhenius temperature dependence of structural relaxation $\tau_\alpha(T)$ [8,9,10], in contrast to super-Arrhenius behavior of $\tau_\alpha(T)$ usually observed in liquids. The sub-Arrhenius behavior corresponds to a strong decrease of the steepness of the $\tau_\alpha(T)$ temperature dependence (fragility) at temperatures close to $T_g$. The fragility index m characterizes the deviation of $\tau_\alpha(T)$ from Arrhenius temperature dependence and is defined as [14,15]:

$$m = \left.\frac{\partial \log[\tau_\alpha(T)]}{\partial\left[\frac{T_g}{T}\right]}\right|_{T=T_g} \quad (1)$$

The fragility index is the lowest, m~20-22, in covalent bonding liquids (e.g. $SiO_2$ and $BeF_2$) and can be higher than m~150 in some polymers [16]. It has been predicted that quantum effects might lead to an unusually low fragility, m<17 [8].

Attempts to search for these quantum effects have been focused first on water [8,9,10], the lightest molecule that exists in a liquid state at ambient conditions. These studies indeed revealed unusually slow temperature variations of $\tau_\alpha(T)$ in the temperature range ~130-150K with m~14, and anomalously large isotope effect on $T_g$ in low-density amorphous (LDA) and vapor deposited (VD) water [9,10]. Both effects are consistent with tunneling contribution to structural relaxation in water. Moreover, its $T_g$~136K is about half of the Debye temperature and thus quantum effects are expected to be non-negligible there [8]. Based on these results, the authors of [8,9,10] suggested that the apparent fragile-to-strong crossover in dynamics of water might be related to the crossover from the classical over-barrier relaxation to tunneling. Unfortunately, it is not possible to verify experimentally this crossover in bulk water due to strong crystallization in the temperature range T~155K – 235K, the so-called 'no-man's land' [17,18,19]. Even more interesting and important question is whether similar effects can be observed in other light molecules with low $T_g$.

This paper presents experimental studies of dynamics of two relatively light molecules with low $T_g$, 3-methyl pentane (3MP) and water. To suppress crystallization, we used confined water. Our studies did not reveal any signs of tunneling in structural relaxation of these liquids, although they show rather low fragility. Confined water show fragility m~18-20, even lower that the least fragile liquids such as $SiO_2$ or $BeF_2$. The isotope effect in confined water appears to be much weaker than in bulk LDA and VD water, but still stronger than in other hydrogen bonding liquids. We show that this effect can be explained by the change in quantum fluctuations (zero-point energy) with H/D substitution. Our model analysis suggests that the suppression of quantum tunneling in comparison to the over-barrier thermally activated hopping in confined water is caused by the reduction of energy barrier for structural relaxation in strong confinement. At the end, we discuss the important molecular parameters that should lead to quantum effects in structural relaxation.

## II. Experimental

The samples of 3-methylpentane with purity higher than 99%, and heavy water, $D_2O$, with 99.9% of D atoms were purchased from Sigma Aldrich. Light water, $H_2O$, of HPLC grade was purchased from Fisher Scientific. All chemicals were used as received without any additional purification.

To suppress crystallization of water we used a silica matrix with 22A pores. It is known that so strong confinement suppresses crystallization of water [20]. High purity commercial grade Davisil® silica gel with 22A in diameter pores was purchased from Sigma-Aldrich. The matrix was dried at temperature 423K under oil-free vacuum of $10^{-10}$ Barr for 2 days to remove any possible water and any other volatile contaminants from the pores. Dry silica matrix was stored in Argon glovebox to ensure its' water-free state. The samples of confined water were prepared by exposing the dry 22A silica gel to the vapors of these liquids inside of an exicator. We estimated the amount of confined liquids by weight. For both $H_2O$ and $D_2O$, silica powder absorbed ~0.4 gram of water per gram of dry powder.

The liquid 3MP was placed between 2 sapphire windows separated by a Teflon o-ring for Brillouin Light Scattering (BLS) measurements. BLS measurements were performed at $90^o$-scattering angle using symmetric scattering geometry that compensates for the refractive index and provides direct estimates of the sound velocities [21]. We used Sandercock tandem Fabri-Perot interferometer and Verdi solid state laser with wavelength $\lambda_{laser}$=532 nm, as described in our earlier publications [22].

All dielectric measurements were carried out using a sealed sample cell with electrodes arranged in parallel plate geometry. For 3MP, a Teflon spacer was used to control the gap between electrodes. For confined liquids no spacer was used as silica gel inherently prevents electrodes from coming into contact. Sealing of the sample cell was done by using o-rings, ensuring that the liquids did not evaporate during measurements. Absence of evaporation was checked by weighing samples before and after the dielectric measurements.

Dielectric measurements of 3MP were performed at the Arizona State University using a Solartron SI-1260 gain-phase analyzer in combination with a Mestec DM-1360 transimpedance amplifier. Temperature was controlled via closed cycle He cryostat with high vacuum sample environment and equipped with Lake Shore LS 340 controller using calibrated diode sensors. Temperature stability during the measurements was within ±0.01 K of the setpoint. Measurements were performed in the frequency range of $2 \cdot 10^{-3} - 1 \cdot 10^{7}$ Hz with the frequency density of 8 points/decade.

Dielectric response of 3MP and confined heavy and light water samples was measured also at the Oak Ridge National Laboratory using a sealed cell with Novocontrol setup equipped with high resolution Alpha impedance analyzer and Quatro temperature control unit. Measurements were performed in the frequency range of $10^{-2} - 10^7$ Hz with the frequency density of 10 points/decade.

### III. Results

The BLS spectrum of 3MP at T=80K (~$T_g$) exhibits both, transverse and longitudinal modes (Fig. 1). The spectrum was fit to a damped harmonic oscillator function to estimate the characteristic frequencies ($\nu_{T,L}$) of the acoustic sound waves. They were used to estimate sound velocities following the equation for the right angle symmetric geometry [22]:

$$C_{T,L} = \frac{\lambda_{laser} \nu_{T,L}}{\sqrt{2}} \qquad (2)$$

We obtained 2.775 km/s for longitudinal and 1.296 km/s for transverse sound velocities. Using these velocities and the equation:

$$\theta_D = \frac{\hbar}{k_B}\omega_D = \frac{\hbar}{k_B}\left(\frac{18\pi\rho}{C_l^{-3} + 2C_t^{-3}}\right)^{1/3} \qquad (3)$$

we estimated the Debye temperature of 3MP $\theta_D \sim 83$ K, assuming density at $T_g$ $\rho(T_g) \sim 1$ g/cm$^3$.

Figure 2 shows the dielectric relaxation spectra of all studied liquids at selected temperatures. The dielectric relaxation process in regular and heavy water show clear peak in the temperature range 250-120K (Fig. 2). The dielectric peak does not have a Debye shape, instead it was symmetrically stretched with characteristic stretching parameters β ≈ 0.6. All the dielectric spectra have been fit to a Havriliak-Negami function plus conductivity. The characteristic relaxation time was taken as the frequency of the maximum in the dielectric loss spectra τ=1/2π$\nu_{max}$. It is known that the dielectric spectra of confined liquids might be affected by the Maxwell-Wagner polarization effects. These effects, however, will not affect the temperature dependence of the obtained relaxation time. So, we did not perform any corrections of the dielectric spectra for the Maxwell-Wagner polarization.

Temperature dependence of the structural relaxation time is presented in the Figure 3. We estimated $T_g$ for all the liquids as the temperature at which extrapolated $\tau_\alpha = 10^3$ s. Fragility index has been estimated as the slope of $\tau_\alpha(T)$ at the lowest temperature range (Eq. 1). The so obtained values of $T_g$ and fragility are presented in the Table 1. The same slope was also used to estimate the apparent activation energy E and the characteristic $\tau_0$ of the apparent Arrhenius behavior at temperature slightly above $T_g$ (Table 1).

## IV. Discussion

### IV.1 Theoretical analysis

Before turning to the analysis of the experimental results let us shortly discuss how structural relaxation and fragility can be affected by quantum effects. In Ref. [8] it was shown that the influence of quantum effects on the temperature dependence of $\tau_\alpha$ can be described by inclusion of zero-point mean squared atomic displacements (MSD) in the usual expression of the elastic theory of the glass transition:

$$\tau_\alpha = \tau_0 \exp\left[\frac{A}{u^2(T)+u_0^2}\right] \qquad (4)$$

where $A$ is a constant, $u^2(T)$ is the thermal part of MSD and $u_0^2$ is zero-point MSD. At sufficiently low T the thermal part becomes smaller than $u_0^2$. At such temperatures the rate of tunneling is comparable to or larger than the rate of thermally activated transitions. If this happens in supercooled state, $T \geq T_g$, the temperature dependence of $\tau_\alpha$ will be unusually weak, apparent activation energy will be decreasing with decreasing temperature and fragility will be unusually small [8]. In some cases even an apparent fragile to strong crossover behavior may occur [8]. To quantify the influence of the quantum effects on fragility $m$, we use Eq. (4) and definition of $m$ in the Eq. (1), resulting in:

$$m = \frac{m_{mol}}{1+u_0^2/u_g^2} \qquad (5)$$

where

$$m_{mol} = m_0 \frac{T_g}{u^2(T)} \frac{du^2(T)}{dT}\Big|_{T=T_g} \qquad (6)$$

is fragility of classical molecular supercooled liquids, $m_{mol}$ ~ 70-90, and $m_0 = \log_{10} \tau_\alpha(T_g)/\tau_0$ ~ 17. In the Debye model of acoustical vibrations $u_T^2 \approx 3k_BT/M\omega_D^2$ and $u_0^2 = 3\hbar/4M\omega_D$ [23], thus Eq.(5) can be rewritten as

$$m \approx \frac{m_{mol}}{1+\hbar\omega_D/4k_BT_g} \quad (7)$$

Using this last equation one can find the dependence of fragility on $T_g$ if one knows how $\omega_D$ scales with $T_g$. The relation between $\omega_D$ and $T_g$ can be found in the following way. In harmonic approximation, $\omega_D$ scales with $T_g$ as it is predicted for an oscillator with a characteristic frequency $\omega_D$ and mass M [24]:

$$M\omega_D^2 u^2(T_g) \sim k_B T_g \quad (8)$$

The Lindemann criterion applied at $T_g$ gives $u^2(T_g) = \gamma^2 a^2$ where $a$ is a characteristic intermolecular distance, $a \propto V_m^{1/3}$, $V_m$ is molecular volume, $M$ is molecular mass and γ ~ 0.12-0.13 is a universal constant. As a result,

$$\omega_D \sim \left(\frac{k_BT_g}{M}\right)^{\frac{1}{2}} \frac{1}{\gamma a} \quad (9)$$

At this point we apply the scaling relation between the molecular mass and $T_g$ in molecular glass-formers that was found in Ref. [25]: $T_g \propto M^\alpha$ where the exponent α ~ 1/2 (the best fit for a few hundreds molecular glass-formers gives α = 0.52 [25]). Assuming also some average density for molecular glass-formers one can find that $a \propto V_m^{1/3} \propto M^{1/3} \propto T_g^{2/3}$. Putting all these scalings into the Eq. (7) gives the following dependence of fragility on $T_g$:

$$m = \frac{m_{mol}}{1+(T_0/T_g)^{13/6}} \quad (10)$$

Here $T_0$ is just a constant, $T_0 = (\hbar/4\gamma k_B^{1/2})^{6/13}(4\pi\rho/3)^{2/13}(T_1^2/M_0)^{5/13}$, where ρ is average density, $M_0$ is atomic mass unit and $T_1$ is defined for molecular glassformers by the relation $T_g = T_1(M/M_0)^{1/2}$. Eq. (6) shows that at high $T_g$, $m$ ~ $m_{mol}$ ~ 70-90. This equation suggests that the low- $T_g$ molecular liquids should have lower fragility due to the contribution of quantum fluctuations to the MSD. Fig.4 presents existing data for fragility and $T_g$ in small molecule organic glass formers. Liquids with $T_g$ higher than ~150-200K exhibit $m$ ~ 70-90, essentially independent of their $T_g$. However, m drops down to below m ~ 50 in liquids with $T_g \leq$ 100-120K. The solid line is the best fit to the Eq. (10), with $m_{mol} = 87$ and $T_0 = 84K$. The constant $T_0$ actually can be estimated if one take the proportionality coefficient connecting $T_g$ and $M^{1/2}$ in molecular glass-formers from Ref. [25] which gives $T_1 = 14.4K$. Taking also some characteristic density for molecular glass-formers, e.g. $\rho = 1.5$ g/cm$^3$, and the Lindemann constant $\gamma = 0.1$, one gets $T_0 = 65K$. Although it is smaller than that found by the fit, it is reasonable taking into account the crude level of the estimate.

## IV.2 Dynamics in 3MP

Having these theoretical predictions, we can now turn to analysis of our experimental data. The presented results (Fig.3, Table 1) demonstrate that none of the studied here liquids exhibit anomalous fragility value. $\tau_\alpha(T)$ in 3MP exhibits normal super-Arrhenius temperature dependence (Fig.3a), although the fragility index m ~ 47 is low for van der Waals liquids that usually have much higher m ~ 70-90 (Fig. 4). The presented analysis of the BLS data reveals that the Debye temperature in 3MP $\theta_D=83K$ is of the order of $T_g$. Thus, according to [8] no significant quantum effects should be expected in this case. This prediction is consistent with the observed behavior of structural relaxation in 3MP (Fig. 3a), and the unusually low for a van der Waals liquid fragility can be explained by the low-$T_g$ of this liquid and relatively high contribution of zero-point oscillations (Eq.10).

## IV.3 Dynamics in confined water

The most intriguing results are the relaxation behavior and the observed isotope effect in confined water (Fig. 3b). $\tau_\alpha(T)$ in both $H_2O$ and $D_2O$ exhibit super-Arrhenius behavior at higher temperature that turns to an Arrhenius-like at lower temperature. This apparent fragile-to-strong crossover is not specific for confined water, and is usual for dynamics of strongly confined liquids [26,27,28]. It has been observed in e.g., confined ethylene glycol [29], salol [30], glycerol [31] and other confined liquids [26,27,28].

It is interesting that confined $H_2O$ exhibits fragility m~18 which differs strongly from the fragility m~14 observed for the bulk LDA and VD water in [9,10]. Also $T_g$ ~115K of the confined $H_2O$ appears to be significantly lower than the value of $T_g$ ~135K reported for LDA and VD water [9,10]. These results clearly indicate a significant difference in behavior of confined water and bulk LDA and VD water. Still, the observed fragility m~18 is the lowest known for any molecular liquids (except LDA and VD water).

Confined $D_2O$ exhibits slightly higher fragility m~19 and higher $T_g$ ~118K. The observed isotope shift in $T_g$, $\Delta T_g$ ~3K, is unusually large for a hydrogen bonding liquid. For example, propanol [32], ethanol [33], glycerol and propylene glycol [9] all exhibit $\Delta T_g$ smaller than 1K upon D/H substitution, although they have relative change of the molecular mass with this substitution similar to that in water. However, the observed isotope shift in $T_g$ for confined water appears much smaller than for bulk LDA and VD water, where anomalously large isotope effect $\Delta T_g$ ~10K has been reported [9]. Thus also the isotope effect appears to be quite different between bulk (LDA and VD) water and confined water (Fig. 5).

The apparent Arrhenius behavior at low temperatures demonstrates another significant difference between behaviors of confined and bulk water (Fig. 5). The apparent activation energy of the confined water E~41 kJ/mol is higher than that observed in the bulk LDA and VD water, E~36 kJ/mol [9,10]. Yet, despite the higher apparent activation energy, the relaxation time appears to be more than 100 times faster in confined water (Fig. 5). This observation is related to much slower $\tau_0$ of the apparent Arrhenius dependence for bulk VD water [9,10]: confined water has reasonable for a light molecule $\tau_0$~$10^{-15}$-$10^{-16}$s, while VD water shows unusually long $\tau_0$~$10^{-10}$-$10^{-11}$s [9, 10] (Fig. 5). This anomalously long $\tau_0$ in bulk VD water has been ascribed to quantum tunneling [9, 10]. Reasonable values for $\tau_0$ and for fragility of confined water (Table 1) suggest no evidence of any tunneling in strongly confined water, in contrast to the behavior of bulk VD and LDA described in Ref. [9].

The absence of tunneling, however, does not mean that quantum effects don't play a role in dynamics of confined water. Detailed analysis of the data (Figs. 3b and Table 1) reveals that the activation energy of the apparent Arrhenius dependence is slightly higher in confined $D_2O$ than in confined $H_2O$. The difference in activation energy is rather small ~1-2 kJ/mol, but it explains why the relaxation time of $D_2O$ is more than 2 times slower than that of $H_2O$ at T~130K (Fig.3b). This difference most probably comes from the so called trivial quantum effects: an effective decrease in the activation energy due to zero-point energy (quantum fluctuations). In quantum mechanics a particle jump over energy barrier not from the bottom of the potential energy well, but from the zero-point energy level $E_0=h\nu_0/2$ (Fig. 6). Dielectric spectroscopy measures reorientation of dipole moments that can be related to a rotational motion of water molecules. In that case we should consider the energy of librational mode of water molecules for estimates of $E_0$. Neutron scattering data indicate librational $h\nu_0 \sim 70\text{-}90$ meV for $H_2O$ that shifts to $h\nu_0 \sim 50\text{-}65$ meV in $D_2O$ [34,35,36]. This shift is consistent with the expected shift $\sim (M_D/M_H)^{1/2} \sim 1.4$ (here $M_D$ and $M_H$ are the mass of deuterium and proton, respectively), because of the motion of H (or D) atoms about oxygen involved in librations. In that case, the change in the zero-point energy upon D/H substitution should be $\sim \Delta h\nu_0/2 \sim 10\text{-}12$ meV $\sim 1$kJ/mol. This value agrees well with the experimentally observed difference in the apparent activation energy of confined $H_2O$ and $D_2O$ at low temperatures (Fig. 5, Table 1), and suggests that the change in zero-point energy is the major reason for the observed difference in dynamics of confined $H_2O$ and $D_2O$, and their relatively strong shift in $T_g$.

Thus although we don't find any signs of tunneling, the quantum effects still play an important role in dynamics of confined water. From that point of view, it is important to compare the behavior of bulk LDA and VD water to the behavior of confined water. Significant difference in their fragility, $T_g$, the activation energy E and $\tau_0$ of the apparent Arrhenius behavior at low T, all can be well explained by the tunneling behavior in bulk water that is suppressed in confined water. The presented in our earlier publication estimates suggest that the energy barrier for the relaxation observed in bulk LDA and VD is $\sim 46$ kJ/mol [10]. So, high energy barrier leads to the probability for a proton (as a part of rotation motion) to tunnel to be higher than the probability to go over the energy barrier at temperatures close to $T_g \sim 136$K [9,10]. It appears that the activation energy for this relaxation is reduced to $\sim 40$ kJ/mol in confined water (Fig.6), the effect known for structural relaxation of other strongly confined liquids [29-33]. As a result an over-barrier relaxation becomes more probable than tunneling in confined water. This leads to the normal value of the apparent $\tau_0$ and lower $T_g$ in confined water (Fig. 4).

Based on these results we propose the following scenario for the observed dielectric relaxation in water. It is dominated by rotational motion of water molecules that involves motion of protons around oxygen atom. In bulk supercooled water (e.g. LDA or VD type) this motion has the energy barrier height (corrected for zero-point energy contribution) $\sim 46$ kJ/mol, and tunneling from higher energy level with $E \sim 36$ kJ/mol in the potential well becomes more probable (for details see [10]). As a result the apparent activation energy of this relaxation appears to be smaller, and the extrapolated $\tau_0$ is too long. This tunneling also leads to a very strong isotope effect upon D/H substitution. Strong confinement of water molecules leads to a significant reduction of the activation energy barrier (Fig.6). The latter may be caused by the reduction in cooperativity in structural rearrangements. As a result the over-barrier relaxation becomes more probable than the tunneling. Yet, significant change in the zero-point energy of quantum fluctuations for libration motions upon D/H substitution leads to a measurable change in the apparent activation energy, slowing down of relaxation and relatively large shift in $T_g$ between confined $H_2O$ and $D_2O$. All these results emphasize a significant difference in dynamics of bulk and confined water, and important role the quantum effects play in water dynamics at low temperatures.

## V.    Conclusions

As a conclusion, the presented analysis of light molecular liquids, bulk 3MP and confined water, does not find any apparent signs of tunneling in their dynamics close to $T_g$. Nevertheless, fragility of these liquids appears to be unusually low for molecular liquids. In the case of water it might be caused in part by strong confinement used to suppress crystallization. However, fragility of bulk 3MP being only m~47 is unusually low for a van der Waals liquid. This observation of low fragility for low $T_g$ liquids is consistent with earlier analysis presented by Qin and McKenna in [37]. We demonstrate that the elastic model of the glass transition, generalized to include zero-point mean squared displacement (quantum fluctuations) predicts such behavior of fragility – it decreases for materials with $T_g \leq 120K$ (Fig.4). The presented theoretical analysis and the experimental data demonstrate that the zero-point vibrations contributing significantly to mean-squared displacements at low temperatures and might lead to slower temperature variations of MSD and related $\log\tau_\alpha(T)$. Describing this in another way, they reduce significantly the energy barriers for relaxation. As a result, they lead to lower fragility than in other glass forming liquids with higher $T_g$.

The role of zero-point quantum fluctuations appears especially important in the case of confined water. The observed isotope effect in dynamics, although much weaker than in the case of bulk LDA or VD water [9,10], can be well explained by the change in the librational zero-point energy. Our results also demonstrate very strong difference in dynamics of bulk LDA or VD water and dynamics of confined water. So, any extrapolations of properties of confined water to that of the bulk water are questionable.

Water is the lightest molecule existing in a liquid state at ambient temperature. This is the major reason it shows strong quantum effects and even signs of tunneling at temperatures ~130-150K [8,9,10]. 3MP is much heavier molecule, and having low $T_g$ ~ 80K is not sufficient for clear appearance of the quantum effects in this liquid. We expect that studies of other molecular systems with mass comparable to water molecule (e.g. methane, ammonia) might reveal many other interesting quantum effects in their structural dynamics close to $T_g$.

**Acknowledgements**

This work was supported by UT-Battelle, LLC. UT team thanks NSF for partial financial support under grant number CHE-1213444.

Table 1. Glass transition temperature $T_g$, fragility *m*, apparent activation energy *E* close to $T_g$ and the respective pre-exponential factor $\tau_0$ for the glass formers studied and discussed in this paper.

|  | $T_g$, K | *m* | E, kJ/mol | $\log\tau_0$, s |
|---|---|---|---|---|
| 3-MP | 78 | 47±1 | 60.4±1 | -38.5±0.7 |
| confined $H_2O$ | 115 | 19.5±1 | 41±1 | -15.7±0.3 |
| confined $D_2O$ | 118 | 19±1 | 43±1 | -16.1±0.1 |
| bulk VD $H_2O$ [9] | 136 | 14±1 | 36±1 | -10.9±0.3 |
| bulk VD $D_2O$ [9] | 146 | 13±1 | 33±1 | -8.8±0.3 |

**Figure captions**

**Figure 1.** Brillouin Light Scattering Spectra of 3MP measured at T=80K, close to its $T_g$.

**Figure 2.** Dielectric relaxation spectra of 3MP (a); confined $H_2O$ (b) and confined $D_2O$ (c) at temperatures 123K, 163K, 203K and 243K.

**Figure 3.** Temperature dependence of the relaxation time estimated from the dielectric loss spectra for 3MP (a), and confined $H_2O$ and $D_2O$ (b). The straight lines show the apparent Arrhenius behavior at T close to $T_g$ that has been used to estimate fragility, activation energy $E$ and $\tau_0$ (see Table 1). Two sets of data for 3MP were measured in Tempe (triangles) and in Oak Ridge (squares).

**Figure 4.** The dependence of fragility on $T_g$ for small molecule organic glass formers. Solid line – fit by Eq. (9) with $T_0$ =84K. Solid squares – data from Ref. [37], stars - data from Ref. [38], up triangle - 3-MP of this work.

**Figure 5.** Low temperature behavior of the dielectric relaxation time for vapor deposited $H_2O$ (red solid squares), vapor deposited $D_2O$ (blue solid circles), confined $H_2O$ (open red triangles) and confined $D_2O$ (open blue diamonds). Solid lines – Arrhenius fit of low-temperature part of relaxation times.

**Figure 6.** Schematic presentation of possible scenario for relaxation in bulk and confined water. Reduction of the energy barrier *E* due to confinement leads to higher probability of the over-barrier jumps than tunneling. However, the role of zero-point quantum fluctuations remains even in confined water, leading to rather strong isotope effect.

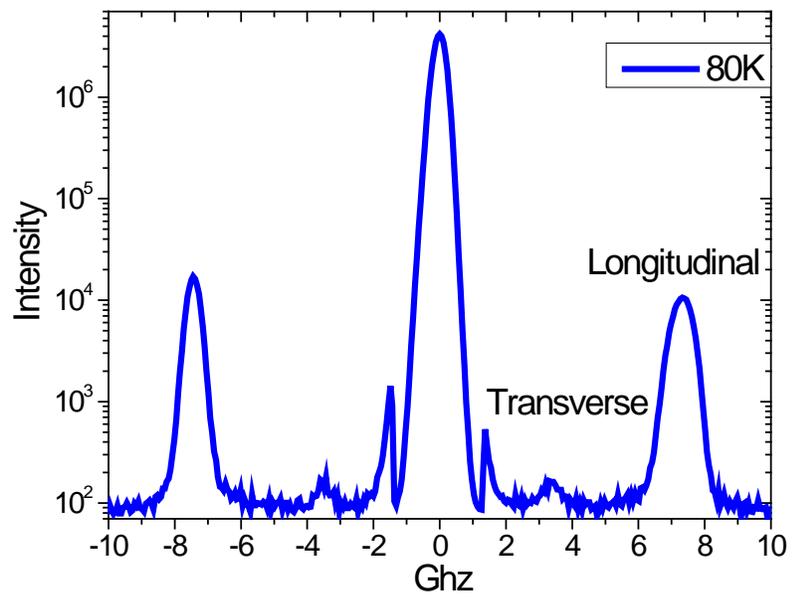

**Figure 1**

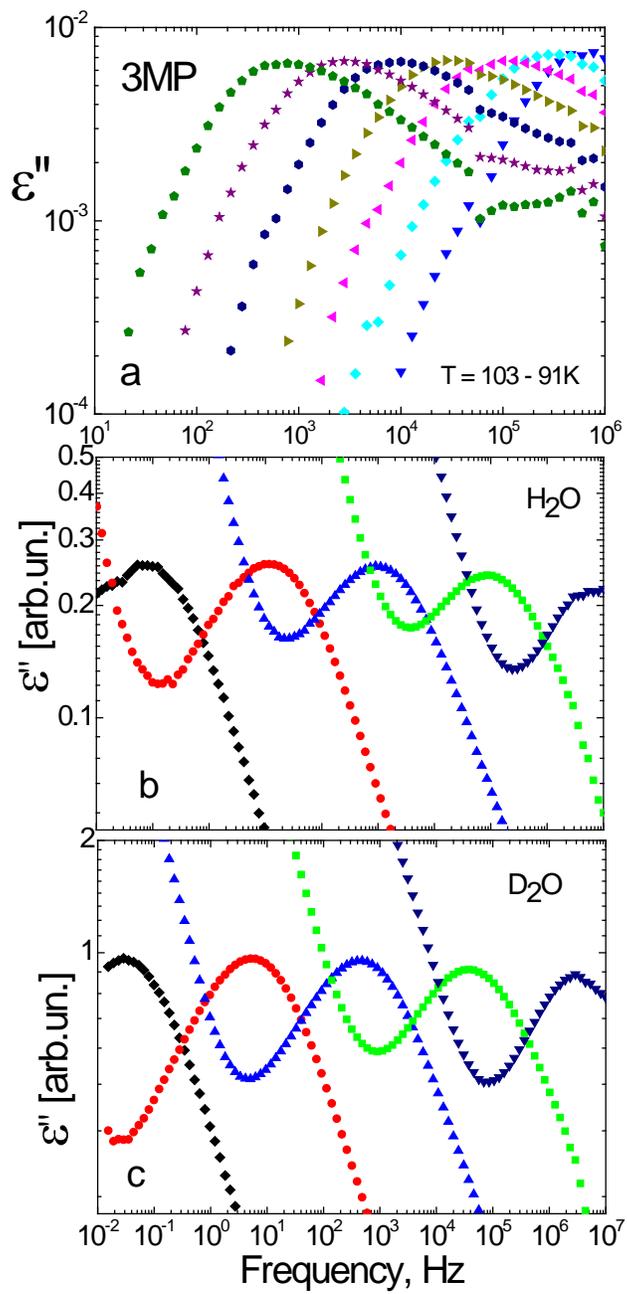

**Figure 2**

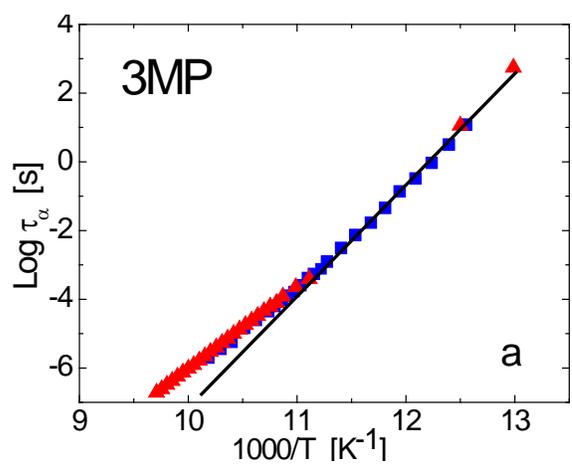

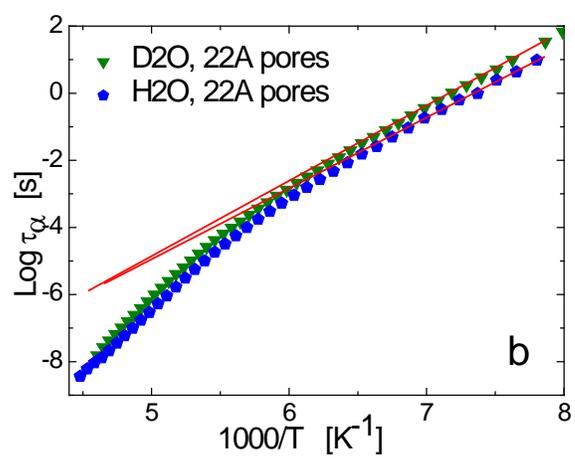

**Figure 3**

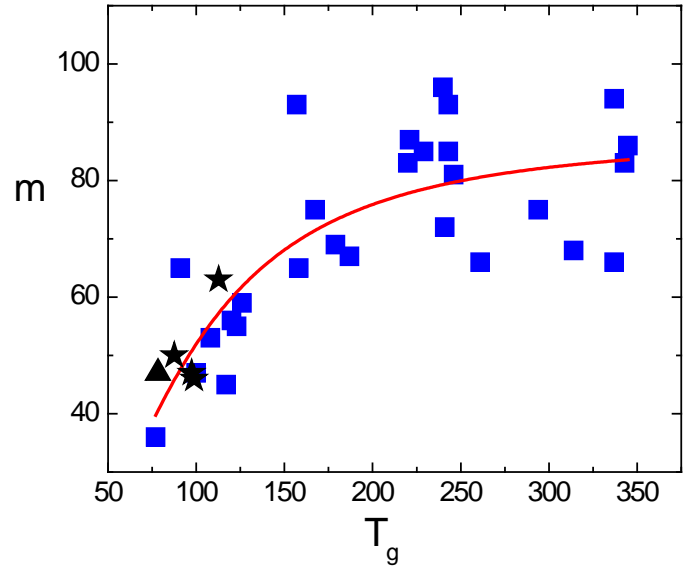

**Figure 4**

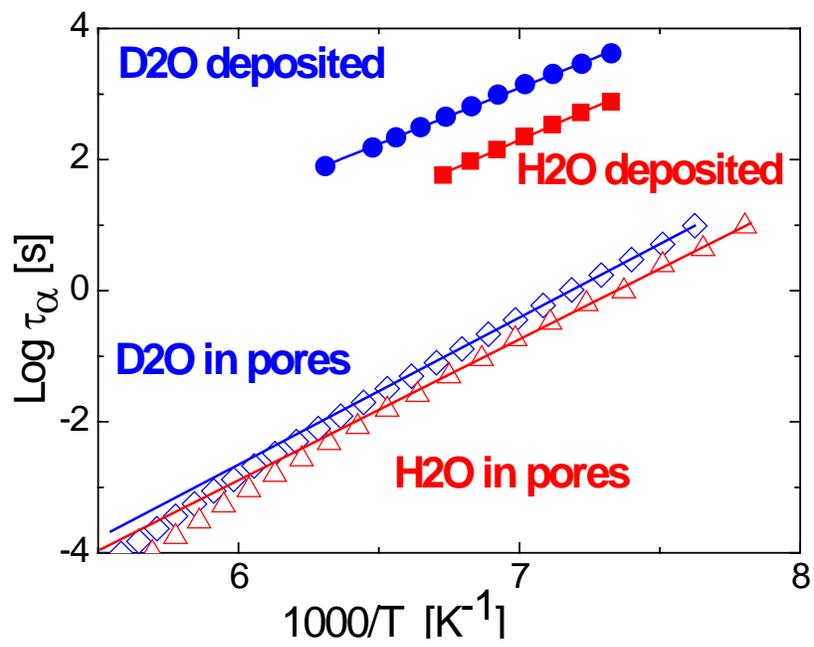

**Figure 5**

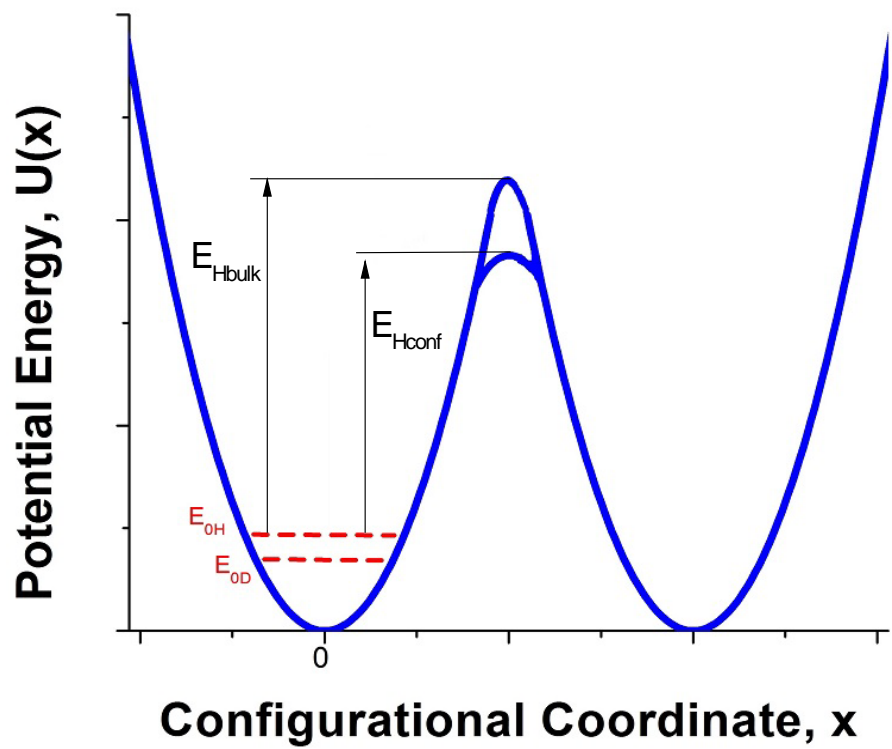

**Figure 6**